\documentclass{emulateapj}

\def\simlt{\mathrel{\rlap{\lower 3pt\hbox{$\sim$}}\raise 2.0pt\hbox{$<$}}}
\def\simgt{\mathrel{\rlap{\lower 3pt\hbox{$\sim$}} \raise 2.0pt\hbox{$>$}}}
\def\lsim{\mathrel{\rlap{\lower 3pt\hbox{$\sim$}}\raise 2.0pt\hbox{$<$}}}
\def\gsim{\mathrel{\rlap{\lower 3pt\hbox{$\sim$}} \raise 2.0pt\hbox{$>$}}}

\def\Msun{{\rm M}_{\odot}}
\def\Zsun{{\rm Z}_{\odot}}

\shortauthors{Salvaterra \& Chincarini}
\shorttitle{The GRB Luminosity Function}

\begin{document}

\title{The Gamma Ray Burst Luminosity Function in the Light of the {\it Swift} 2--year Data}

\author{
R.~Salvaterra,\altaffilmark{1}
G.~Chincarini,\altaffilmark{1,2}
}
\altaffiltext{1}{Dipertimento di Fisica G.~Occhialini, Universita degli Studi di Milano
Bicocca, Piazza della Scienza 3, I-20126 Milano, Italy, salvaterra@mib.infn.it}
\altaffiltext{2}{INAF, Osservatorio Astronomico di Brera, via E. Bianchi 46, I-23807 Merate (LC), Italy}

\begin{abstract}
We compute the luminosity function (LF) and the formation rate of long gamma 
ray bursts (GRBs) by fitting the observed differential peak flux distribution
obtained by the {\it BATSE} satellite in three different
scenarios:  i) GRBs follow the cosmic star formation and their LF is constant 
in time; ii) GRBs follow the cosmic star formation but the LF varies with
redshift; iii) GRBs form preferentially in low--metallicity
environments. We find that the differential peak flux number counts 
obtained by {\it BATSE} and by {\it Swift} can be reproduced using 
the same LF and GRB formation rate, indicating that 
the two satellites are observing the same GRB population. 
We then check the resulting redshift distributions in the light of  
{\it Swift} 2--year data, focusing in 
particular on the relatively large sample of GRBs detected 
at $z>2.5$. We show that models in which GRBs trace the cosmic star 
formation and are described by a constant LF are ruled out by the number 
of high--$z$ {\it Swift} detections. This conclusion does not depend on 
the redshift distribution of bursts that lack
of optical identification, nor on the existence of a decline in star formation 
rate at $z>2$, nor on the adopted faint--end of the GRB LF. {\it Swift}
observations can be explained by assuming that the LF varies with redshift 
and/or that GRB formation is limited to low--metallicity environments.
\end{abstract}


\section{Introduction}

Gamma Ray Bursts (GRBs) are powerful flashes of high--energy photons occurring
at an average rate of a few per day throughout the universe. Even
though they are highly transient events very hard to localize, they are
so bright that they can be detected up to very high redshift (the current
record is $z=6.29$). The energy source of a GRB is believed
to be associated to the collapse of the core of a massive star in the case
of long--duration GRBs, and due to merger-- or accretion--induced collapse 
for the short--hard class of GRBs (see M\'esz\'aros 2006 for a recent review). 
In this paper, we limit our analysis to the class of long--duration GRBs.

One of the main goals of the 
{\it Swift}  satellite (Gehrels et al. 2004) is to trackle the key issue 
of the GRB luminosity function (LF). Unfortunately, although the number of 
GRBs with good redshift
determination has been largely increased by {\it Swift}, 
the sample is still too poor (and bias dominated) to allow a direct 
measurement of the LF. 
Many studies (e.g. Lamb \& Reichart 2000; Porciani \& Madau 2001 (PM01); Schmidt 
2001, Choudhury \& Srianand 2002; Firmani et al. 2004; Guetta, Piran \& Waxman 
2005; Natarajan et al.2005; Daigne, Rossi \& Mochkovitch 2006)  
tried to constrain the GRB LF under the assumption that GRBs trace the 
observed star formation rate, as suggested by the association of long GRBs to 
the death of massive stars.
Following these works and assuming the most 
recent star formation rate determination, we 
derive the LF and formation rate of GRBs by fitting the observed 
{\it BATSE} differential peak flux distribution in three different 
scenarios: i) GRBs 
follow the cosmic star formation and have a constant LF; ii) the
GRB  LF varies with redshift; iii) GRBs form in low--metallicity
environments. We check the results against the 2--year {\it Swift} data, 
focusing in particular on the large sample of high redshift ($z>2.5$) GRBs 
detected by this instrument.


\section{Basic Equations}

The observed photon flux, $P$, in the energy band 
$E_{\rm min}<E<E_{\rm max}$, emitted by an isotropically radiating source 
at redshift $z$ is

\begin{equation}
P=\frac{(1+z)\int^{(1+z)E_{\rm max}}_{(1+z)E_{\rm min}} S(E) dE}{4\pi d_L^2(z)},
\end{equation}

\noindent
where $S(E)$ is the differential rest--frame photon luminosity of the source, 
and $d_L(z)$ is the luminosity distance. 
To describe the typical burst spectrum we adopt the
functional form proposed by Band et al. (1993), i.e. a broken power--law
with a low--energy spectral index $\alpha$, a high--energy spectral index
$\beta$, and a break energy $E_b$. In this work, we take $\alpha=-1$ and
$\beta=-2.25$ (Preece et al. 2000), and $E_b=511$ keV (PM01).
Moreover, it is customary to define an isotropic equivalent intrinsic burst 
luminosity  in the energy band 30-2000 keV as 
$L=\int^{2000\rm{keV}}_{30\rm{keV}} E S(E)dE$. Given a normalized GRB LF, 
$\phi(L)$, and the detector efficiency, $\epsilon(P)$, the observed rate of 
bursts with peak flux between $P_1$ and $P_2$ is

\begin{eqnarray}
\frac{dN}{dt}(P_1<P<P_2)&=&\int_0^{\infty} dz \frac{dV(z)}{dz} 
\frac{\Delta \Omega_s}{4\pi} \frac{\Psi_{\rm GRB}(z)}{1+z} \nonumber \\
& \times & \int^{L(P_2,z)}_{L(P_1,z)} dL^\prime \phi(L^\prime)\epsilon(P),
\end{eqnarray}

\noindent
where $dV(z)/dz=4\pi c d_L^2(z)/[H(z)(1+z)^2]$ is the comoving volume 
element\footnote{We adopted the 'concordance' model values for the
cosmological parameters: $h=0.7$, $\Omega_m=0.3$, and $\Omega_\Lambda=0.7$.},
and $H(z)=H_0 [\Omega_M (1+z)^3+\Omega_\Lambda+(1-\Omega_M-\Omega_\Lambda)(1+z)^2]^{1/2}$.
$\Delta \Omega_s$ is the solid angle covered on the sky by the survey,
and the factor $(1+z)^{-1}$ accounts for cosmological time dilation. 
Finally, $\Psi_{\rm GRB}(z)$ is the comoving burst formation rate. In this 
work, we assume that the GRB LF is described by

\begin{equation}
\phi(L) \propto \left(\frac{L}{L_{\rm cut}}\right)^{-\xi} \exp \left(-\frac{L_{\rm cut}}{L}\right).
\end{equation}

\section{Models}

We consider three different scenarios. In the first one, 
the GRB formation rate is proportional to the cosmic star formation rate 
(SFR), $\Psi_\star(z)$, i.e. $\Psi_{\rm GRB}(z)=k_{\rm GRB} \Psi_\star (z)$, 
and the LF does not evolve with redshift, i.e.
$L_{\rm cut}={\rm const}=L_0$. The factor $k_{GRB}$ gives the number of 
GRBs formed per solar mass in stars and has units of $\Msun^{-1}$.
$\Psi_\star(z)$ (in units of $\Msun$ Mpc$^{-3}$ yr$^{-1}$) is 
commonly parameterized with the form proposed by Cole et al. (2001) as

\begin{equation}
\Psi_\star(z)=\frac{(a_1+a_2z)h}{1+(z/a_3)^{a_4}}. 
\end{equation}

\noindent
Recently, Hopkins \& Beacom (2006) have provided the values of the coefficients
$a$ by fitting the available UV and far--infrared measurements for $z<6$, 
corrected for dust obscuration. In this paper, we adopt their best fit 
parameters: $a_1=0.017$, $a_2=0.13$, and $a_3=3.3$ (Hopkins \& Beacom 2006). 
The value of $a_4=4.3$ is
taken to be slightly lower than the original one in order to match the 
decline of the SFR with $(1+z)^{-3.3}$ at $z\gsim 5$ suggested by recent
deep--field data (see  Stark et al. 2006 and references therein).

In the second scenario, while the GRB formation rate is still proportional to 
the observed SFR, the cut--off luminosity in the GRB LF increases with 
redshift as $L_{\rm cut}=L_0 (1+z)^\delta$. Lloyd--Ronning, Fryer \&
Ramirez--Ruiz (2002), using GRB redshifts and luminosities
derived from the luminosity--variability relationship, found that the data 
imply $\delta\simeq 1.4\pm0.5$, and we adopt this as fiducial value.

Finally, we consider a case in which GRBs form only in environments with 
metallicity below a given threshold (no evolution in the LF is considered). 
In fact, some theoretical models (see M\'esz\'aros 2006 and reference therein) 
require that GRB progenitors should have metallicity $\lsim 0.1\;\Zsun$.
Observations of GRB host galaxies (see Savaglio 2006 and
reference therein) seems in agreement with this prescription, showing that GRB 
preferentially originates in low--metallicity regions. 
Langer \& Norman (2006) have quantified the amount of star formation at a given
metallicity, using a recent determination of the stellar mass function 
(Panter et al. 2004) and the observed mass--metallicity correlation 
(Savaglio et al. 2005). Adopting the metallicity redshift evolution derived
from emission line studies (Kewley \& Kobulnicky 2005), the fractional mass
density belonging to metallicity below a given threshold, $Z_{th}$, can be 
computed as 

\begin{equation}
\Sigma(z)=\frac{\hat{\Gamma}(0.84,(Z_{th}/\Zsun)^2 10^{0.3z})}{\Gamma (0.84)},
\end{equation}

\noindent
where $\hat{\Gamma}$ ($\Gamma$) are the incomplete (complete) gamma
function, and $\Gamma(0.84)\simeq 1.122$. The GRB formation rate is then 
given by $\Psi_{\rm GRB}(z)= k_{\rm GRB}\Sigma(z)\Psi_\star(z)$.  
The main effect of this convolution is that the GRB
formation rate peaks at higher redshift with respect to the cosmic SFR. 
We adopt $Z_{th}=0.1\;\Zsun$ as fiducial value, and, in this case, the GRB 
formation peaks at $z\sim 3.5$.

\section{GRB number counts}

The free parameters in our model are the GRB formation efficiency 
$k_{\rm GRB}$, the cut--off luminosity at $z=0$, $L_0$, and the power index,
$\xi$, of the GRB LF function. Following PM01, we optimized
the value of these parameters by $\chi^2$ minimization over the observed
differential number counts in the 50--300 keV band of {\it BATSE}. 
We use the off--line
{\it BATSE} sample of Kommers et al. (2000), which includes 1998 archival
(``triggered'' plus ``non--triggered'') bursts, and for which the detector
efficiency is well described by the function 
$\epsilon(P)=0.5[(1+{\rm erf}(-4.801+29.868 P)]$ (Kommers et al. 2000). 
We report the best--fit parameters for our fiducial models in 
Table~\ref{tab:fit}. 
In the last column, we give the reduced $\chi^2$ for the
best--fitting model, showing that it is always possible to find a good 
agreement with the data\footnote{Note that strong covariance on $L_0$ and 
$\xi$ is observed in the parameter space surrounding the best--fit parameters
(see also PM01)}.
Note that for the metallicity evolution scenario a higher GRB formation 
efficiency is required, since GRBs form only in a (small) fraction 
of star forming galaxies.

\begin{figure}
\epsscale{1.0}
\plotone{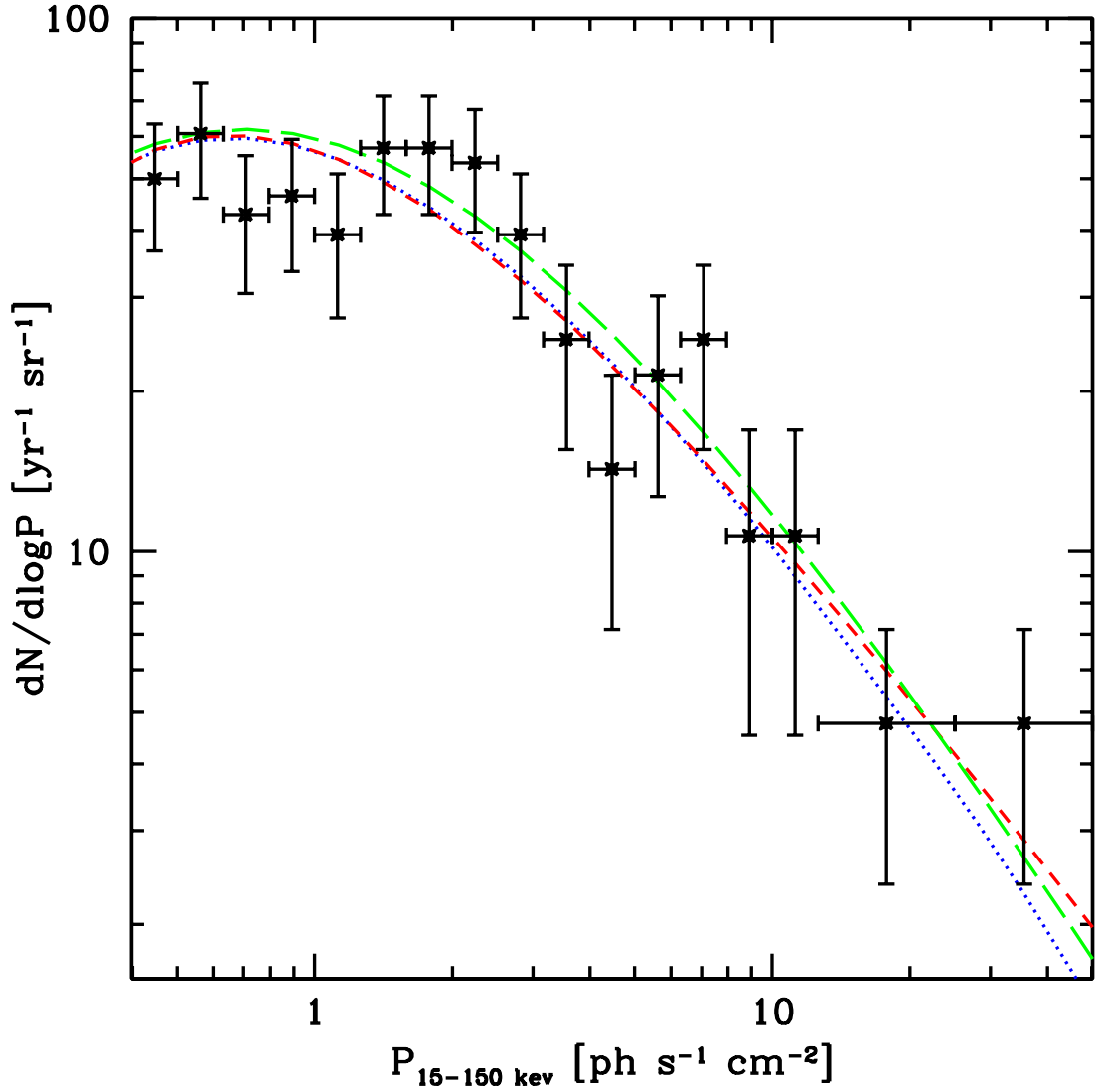}
\caption{Differential number counts for {\it Swift} in the 15--150 keV band
as a function of the observed photon flux $P$. The points show the observed
counts and their Poisson uncertainties (horizontal error bars denote bin
size). Dotted lines refers to the model without evolution, short--dashed line 
to the luminosity evolution model ($\delta=1.4$), and 
long--dashed line to the model with the metallicity threshold for GRB formation
($Z_{th}=0.1\;\Zsun$). A field of view of 1.4 sr for {\it Swift}/BAT is 
adopted.}
\label{fig:swift}
\end{figure}

We can now use the best--fit parameters to compute the expected differential 
peak flux distribution of GRBs in the 15--150 keV band of the Burst Alert 
Telescope (BAT) instrument onboard of {\it Swift}. The results are plotted 
in Figure~\ref{fig:swift} and
compared with the observed {\it Swift}/BAT data points. 
All models show a good agreement with the data without the need of any change
of the GRB LF and formation efficiency, indicating that 
{\it BATSE} and {\it Swift} are observing essentially the same 
population of GRBs. This conclusion is rather insensitive to
$20$\% variations of the adopted GRB spectrum parameters, i.e. for
the large majority of burst spectra (Kaneko et al. 2006).

\section{GRB redshift distribution}

Our model allows us to compute the expected redshift distribution of GRBs
detected by {\it Swift}. We decide to avoid the comparison between model 
results and
the overall observed distribution of bursts with known redshift, since 
this procedure implicitly assumes that the observed sample of GRBs with 
redshift determination is representative of all detected sources. 
Moreover, important information are missed by this kind of analysis: for 
example, that many bright GRBs are identified at high redshift.
So, we try to answer this simple question: {\it is the redshift
distribution consistent with the number of {\it Swift} detections at 
$z>2.5$ and $z>3.5$?}

\begin{figure}
\epsscale{1.0}
\plotone{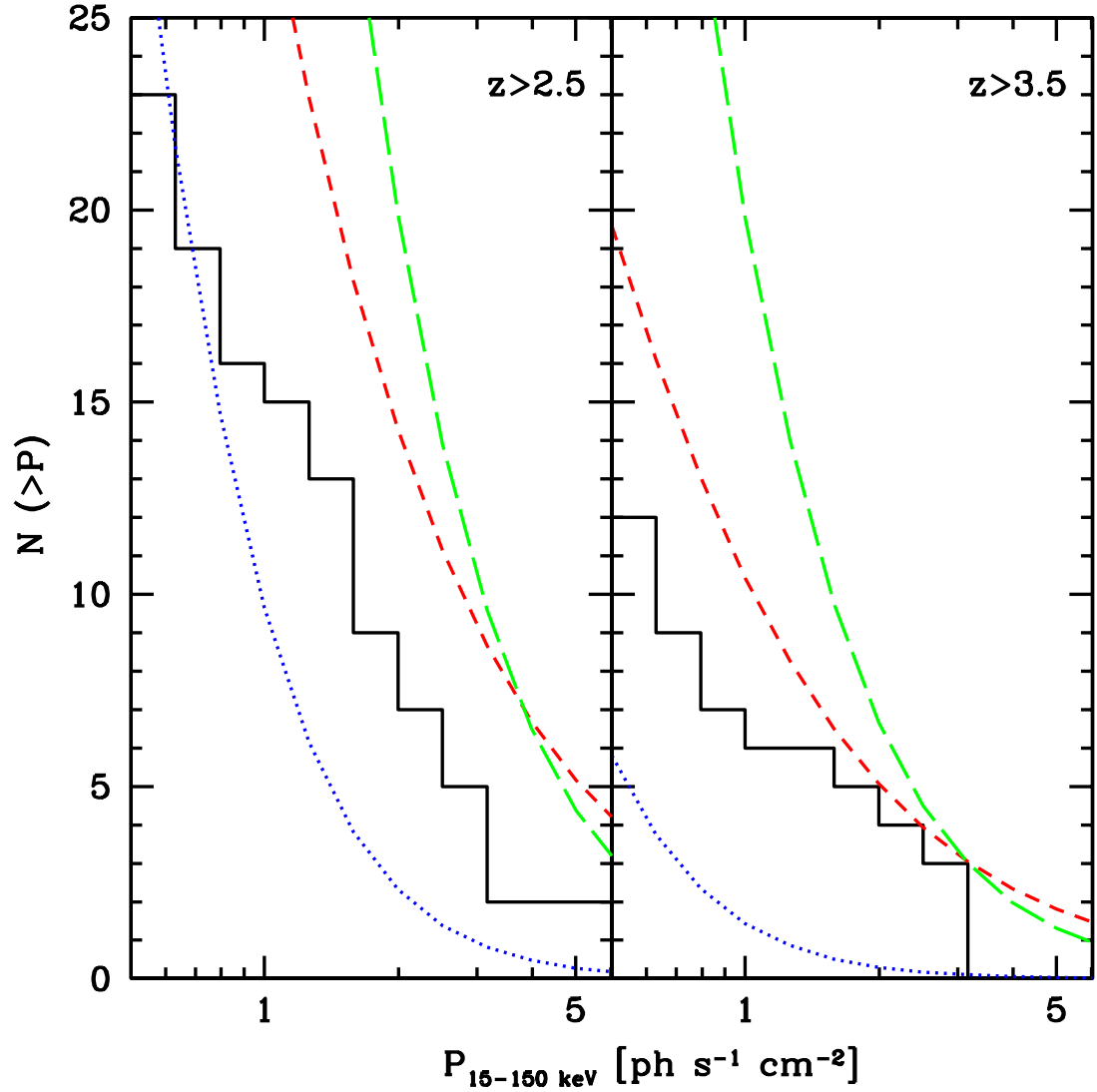}
\caption{Cumulative number of high redshift GRBs at $z>2.5$ (right panel) and
at $z>3.5$ (left panel) as a function of the observed photon flux $P$ in the 
15--150 keV band. The number of sources detected in the two years of 
{\it Swift} mission is shown as solid histogram, whereas model results are 
shown with lines as in the previous figure. Note that the observed detections 
are lower limits, since many high--$z$ GRBs can be missed by optical 
follow--up searches. A field of view of 1.4 sr for {\it Swift}/BAT is 
adopted.}
\label{fig:zgt}
\end{figure}

The cumulative number of GRBs, identified during the two years of the 
{\it Swift} mission  at $z>2.5$ (left panel) and $z>3.5$ (right 
panel), is plotted in Figure~\ref{fig:zgt} toghether with model predictions. Note
that {\it Swift} detections are to be considered as a strong lower limit, 
since many high--$z$ bursts can be missed by optical follow--up searches.

The model with no LF evolution clearly underestimates the number of high redshift
GRB detections at any photon flux and no bright GRBs are predicted for $z>3.5$.
We checked that variations of the shape of the SFR do not affect 
this result: even assuming a constant SFR at $z\gsim 2$, the model predictions 
do not change significantly. In fact, for relatively bright GRBs, the rapid 
decline in the LF strongly hampers the detection of high redshift bursts. 
Furthermore, our analysis does not depend on the faint--end of the LF:
increasing the population of faint GRBs would decrease the number of
high--$z$ detections, strengthening our conclusion. So, models 
in which GRBs trace the cosmic SFR and are described by a 
constant LF, are ruled out by the large sample of high--$z$ {\it Swift} 
GRBs.

The number of high--$z$ {\it Swift} identifications can be justified 
assuming that the LF varies with redshift. In this case, high--$z$ GRBs 
are typically brighter than low--$z$ ones, so that are much easely detected.
Assuming that the luminosity increases as  $(1+z)^{1.4}$, we find 
many sources at $z>2.5$, but the model is barely consistent with the number 
of bright GRBs at $z>3.5$. Since some high--$z$ sources can be 
missed by optical follow--up searches, an even stronger evolution might be 
required to explain the data. 

Finally, we consider the possibility that GRB formation is restricted to 
low--metallicity environments. In this case, the peak of the GRB 
formation is shifted towards higher redshift, so that
the probability of high--$z$ detections increases.
Assuming $Z_{th}=0.1\;\Zsun$, {\it Swift} identification are exceeded
both at $z>2.5$ and $z>3.5$ without requiring any evolution in the LF. 
Thus, the model is consistent with a fraction of high redshift bursts missed 
by optical follow--up searches. Increasing the threshold metallicity will 
decrease the number of sources at high--$z$: for $Z_{th}\sim 0.4\;\Zsun$ the 
model becomes inconsistent with the number of observed GRBs at $z>3.5$.
Higher threshold values would require evolution of the GRB
luminosity and/or a more gentle decline of the SFR at high redshift. 

In conclusion, the existence of a large sample of bursts at
$z>2.5$ in the {\it Swift} 2-year data imply that 
GRBs have experienced some kind of evolution, being more luminous or
more common in the past.

\section{GRB rate at redshift larger than six}

The discovery of GRB050904 at $z=6.29$ (Antonelli et al. 2005; 
Tagliaferri et al. 2005; Kawai et al. 2006) during the first year of the 
{\it Swift} mission has strengthened 
the idea that many GRBs should be observed out to very high redshift 
(e.g. Natarajan et al. 2005; Bromm \& Loeb 2006; Daigne et al. 2006).  
Unfortunately, no other source at $z\gsim 6$ has been detected in the second 
year of observations.

\begin{figure}
\epsscale{1.0}
\plotone{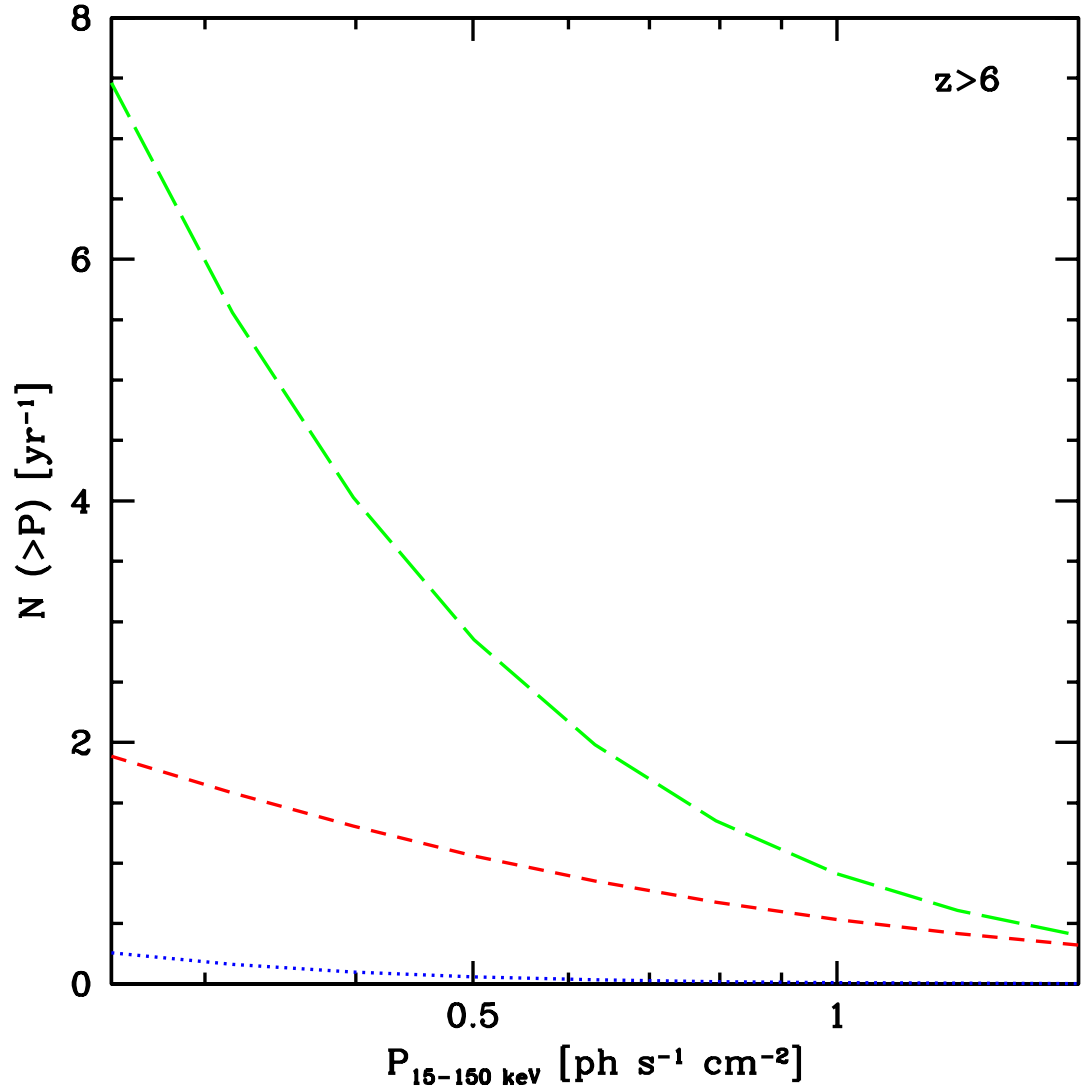}
\caption{Cumulative rate of $z\gsim 6$ GRBs detectable by {\it Swift} as a 
function of the photon flux $P$. A field of view of 1.4 sr for 
{\it Swift}/BAT is adopted. Lines as in Figure~\ref{fig:swift}.}
\label{fig:zgt6}
\end{figure}

In Figure~\ref{fig:zgt6}, we plot the {\it Swift} detection rate expected for
the three scenarios here considered. Models without evolution predict almost
no sources to be detected at very high redshift. If luminosity evolution
($\delta=1.4$) is allowed, $\sim 2$ bursts/yr should lie above 
$z\sim 6$ for $P>0.2$ ph cm$^{-2}$ s$^{-1}$, whereas, in the metallicity 
evolution scenario ($Z_{th}=0.1\;\Zsun$), we expect $\sim 8$ GRBs/yr, 
one or two being at $z\gsim 8$. 

The detection rate are found to decrease rapidly with increasing peak fluxes. 
Indeed, it is interesting to note that GRB050904 was relatively 
bright, being its observed photon flux $P=0.658$ 
ph cm$^{-2}$ s$^{-1}$. At this limit, only $\sim 1$ (2) bursts/yr 
would be at $z\gsim 6$, if luminosity (metallicity) evolution is assumed. 
Thus, the lack of very high redshift identification in the 2nd year of
the {\it Swift} mission might be due to 
practical difficulties in the optical follow--up of faint GRBs. 
In fact, no GRB with observed photon fluxes below $0.5$ ph cm$^{-2}$ s$^{-1}$
has UVOT detection  and only in a couple of cases a reliable redshift 
determination was possible.
So, the identification of just one burst at $z\gsim6$ in two years of 
{\it Swift} mission is not very surprising. On the contrary, the discovery 
of GRB050904 may suggest that the {\it Swift} follow--up procedure is 
working very well, at least for relatively bright bursts.

\section{Conclusions}

We have computed the luminosity function and the formation rate of long GRBs 
by fitting the {\it BATSE} differential peak flux number counts in three 
different scenarios:  i) GRBs follow the cosmic star 
formation and have a redshift--independent LF; ii) the GRB LF varies with 
redshift; iii) GRBs are associated with star 
formation in low--metallicity environments. In all cases, it is possible to 
obtain a good fit to the data by adjusting the model free parameters.
Moreover, using the same LF and formation rate, it is possible to 
reproduce both {\it BATSE} and {\it Swift} differential counts, showing 
that the two satellites are observing the same GRB population.

We have then computed the expected burst redshift distribution, testing the 
results against the number of high redshift GRBs, detected during the two 
years of the {\it Swift} mission. We find that models where GRBs trace the 
SFR and are described by a constant LF largely underestimate the number
high--$z$ GRBs detected by {\it Swift}. This conclusion
does not depend on the redshift distribution of burst lacking
of optical identification, nor on the existence of a decline in the SFR at
$z>2$,  nor on the adopted faint--end of the LF.
Alternatively, we find that the observed number of high--$z$
detection can be justified by assuming that the GRB luminosity increases with 
redshift and/or that GRBs preferentially form in low--metallicity environments. 

Finally, we have estimated the detection rate of bursts at very high 
redshift. We find that $\sim 2$ (8) GRBs/yr should be observed 
at $z\gsim 6$, if luminosity (metallicity) evolution is assumed.
The majority of these sources is faint and may be missed in optical 
follow--up searches, but $\sim 1$ (3) GRB/yr should be relatively
bright, with an observed photon flux in excees to 0.5 ph cm$^{-2}$ s$^{-1}$.

\clearpage

\begin{deluxetable}{lcccc}
\tablecolumns{5}
\tablewidth{0pt}
\tablecaption{Best fit parameters for different models. Errors are at 1$\sigma$
level.}
\tablehead{\colhead{Model} & \colhead{$k_{\rm GRB}/(10^{-8}\Msun^{-1})$}  & \colhead{$L_0/(10^{51} {\rm \;erg\;s}^{-1})$} & \colhead{$\xi$} & \colhead{$\chi_r^2$}}
\startdata
no evolution & $1.14\pm0.07$ & $9.54\pm4.55$ & $3.54\pm0.78$ & 0.83 \\ 
luminosity evolution ($\delta=1.4$) & $1.05\pm0.05$ & $0.77\pm0.13$ & $2.19\pm0.95$ & 0.80 \\ 
metallicity evolution ($Z_{th}=0.1\;\Zsun$) & $10.0\pm0.5$ & $16.7\pm5.7$ & $2.94\pm0.34$ & 0.84 \\ 
\enddata
\label{tab:fit}
\end{deluxetable}

\clearpage

\clearpage

\end{document}